\documentclass[10pt,conference]{IEEEtran}\IEEEoverridecommandlockouts
\usepackage{epsfig,epstopdf,graphics,subfigure,psfrag,amsmath,cases}
\usepackage{latexsym,amssymb,amsmath,epsfig,subfigure,algorithm,amsthm}
\usepackage{algorithmic}
\usepackage{color}
\usepackage{url}
\usepackage{scrtime}
\usepackage{bm}
\usepackage{graphicx}

\usepackage{subfigure}
\usepackage{bbding}
\usepackage{multicol}
\usepackage{graphics}
\usepackage{amsthm}
\usepackage{cite}
\usepackage{array}
\usepackage{multirow}
\title{Connecting Spatially Coupled LDPC Code Chains for Bit-Interleaved Coded Modulation}
\author{{Yihuan~Liao,
		Min~Qiu,
		and~Jinhong~Yuan}
	\IEEEauthorblockA{\\School of Electrical Engineering and Telecommunications, University of New South Wales, Sydney, NSW, Australia}\thanks{The work was partially supported by the Australian Research Council (ARC) Discovery Projects under Grant DP190101363 and by the ARC Linkage Project under Grant LP170101196.}}

{}

\usepackage{mathtools}

\newcommand{\abs}[1]{\lvert#1\rvert}

\begin{document}
\maketitle

\begin{abstract}
	This paper investigates the design of spatially coupled low-density parity-check (SC-LDPC) codes constructed from connected-chain ensembles for bit-interleaved coded modulation (BICM) schemes. For short coupling lengths, connecting multiple SC-LDPC chains can improve decoding performance over single-chains and impose structured unequal error protection (UEP). A joint design of connected-chain ensembles and bit mapping to further exploit the UEP from codes and high-order modulations is proposed. Numerical results demonstrate the superiority of the proposed design over existing connected-chain ensembles and over single-chain ensembles with existing bit mapping design.
\end{abstract}

\section{Introduction}
\IEEEPARstart{S}{patially} coupled low-density parity-check (SC-LDPC) codes have received a lot of attention in both academia and industry \cite{782171, 5613891,5571910,5695130,6589171,7152893,8357804}. The idea of coupling a sequence of identical LDPC block code (LDPC-BC) graphs into a chain was first proposed in \cite{782171}. Additionally, by using iterative belief propagation (BP) decoding, the reliable information can propagate from the boundaries to the middle of the chain. Furthermore, the BP decoding threshold of SC-LDPC codes achieves the maximum a posteriori (MAP) threshold of the individual uncoupled codes. This is a phenomenon known as \textit{threshold saturation} \cite{5613891,5571910,5695130,6589171}. Threshold saturation requires the coupling lengths of SC-LDPC codes to be very large. Whereas, for small coupling length, there exists a noticeable gap between the BP decoding threshold and the MAP threshold \cite{8718022}. 

Recently, connected-chain SC-LDPC codes were introduced by extending the spatial graph coupling phenomenon from coupling individual LDPC-BC graphs in a chain to more general coupled structures \cite{6364482,6181771,8718022}. Specifically, several individual SC-LDPC code chains, referred to as the \textit{single-chain ensembles}, are connected to form a \textit{connected-chain ensemble}. These connected-chain SC-LDPC codes have a decoding performance improvement of about $0.4$ dB over single-chain SC-LDPC codes for binary phase-shift keying (BPSK) modulation with the same codeword length and code rate \cite{8718022}. Whereas, the authors in \cite{6955088,8003438} proposed a continuous transmission by connecting multiple chains in a layered format, where encoding, transmission, and decoding are performed in a continuous fashion.

In modern communication systems, coded modulation (CM) is indispensable for achieving high spectral efficiency. Bit-interleaved coded modulation (BICM), as a pragmatic approach in CM, has been extensively investigated for many wireless and optical communication systems \cite{669123, bicm_wireless,5205587,5729846,9373632}. It is well known that a properly designed bit mapping, determines the mapping of coded bits to bit-channels, provides non-negligible performance improvement in high-order modulation schemes with unequal error protection (UEP) property. The bit mapping design for single-chain SC-LDPC coded BICM schemes has been investigated in \cite{6883627} by using density evolution \cite{910577}. Furthermore, the authors in \cite{Hager:15} design the bit mapping using the protograph-based extrinsic information transfer chart (PEXIT) \cite{PEXIT_chart}. However, when it comes to small coupling lengths, the performance of single-chain SC-LDPC coded BICM noticeably differs from the capacity limit. Connected-chain SC-LDPC codes perform better than their single-chain counterparts at small coupling lengths for BPSK modulation. However, the design of connected-chain SC-LDPC codes for high-order modulations has not been reported in the literature yet.  

The design of connected-chain SC-LDPC codes requies determining the type and the number of single chains to be connected, the coupling length, the choice of connect-positions, and the structure of connection. Therefore, the bit mapping design needs to take into account all the these factors. This paper investigates the design of connected-chain ensembles and the bit mapping for BICM systems. Specifically, a new method is proposed to connect multiple single-chains, introducing UEP among VNs within each chain-position connected to other chains. We consider the mapping of coded bits to bit-channels in BICM using connected-chain SC-LDPC codes constructed from the proposed method. Then, we develop a differential evolution \cite{diffevo} based optimization algorithm to jointly optimize the connect-positions and the bit mapping to improve the decoding threshold computed via density evolution. Numerical results show that our jointly designed connected-chain SC-LDPC code and the bit mapper outperform the existing connected-chain SC-LDPC codes from \cite{6364482,6181771,8718022,6955088,8003438} with the bit mapper design in \cite{6883627, Hager:15}. 

\section{System Model} \label{sec:system_model}
\begin{figure}[h]
	\centering
	\includegraphics[width=0.45\textwidth]{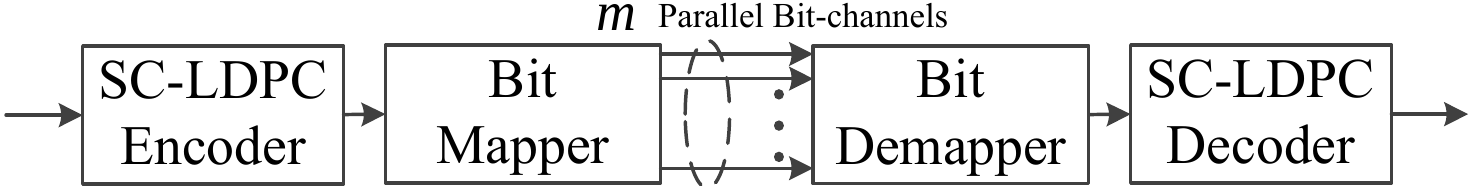}
	\caption{Block diagram of SC-LDPC coded BICM system.~~~~~~}
	\label{fig:BICM}
\end{figure}

In this paper, we consider SC-LDPC coded BICM schemes over an additive white Gaussian noise (AWGN) channel. The block diagram of an SC-LDPC coded BICM system is shown in Fig. \ref{fig:BICM}. We consider the AWGN channel $y=x+z$, where $x\in\chi$ is the channel input from a constellation $\chi$ and $z \sim \mathcal{CN}(0,1)$. Each symbol in a constellation $\chi$ is mapped to a unique binary string of length $m=\log_2\abs{\chi}$. These $m$ bits are transmitted in $m$ parallel bit-channels with capacity $C_{i}, i\in\{0,1,\cdots,m-1\}$.

\begin{figure}[h]
	\vspace*{-4mm}
	\centering
	\includegraphics[width=0.45\textwidth]{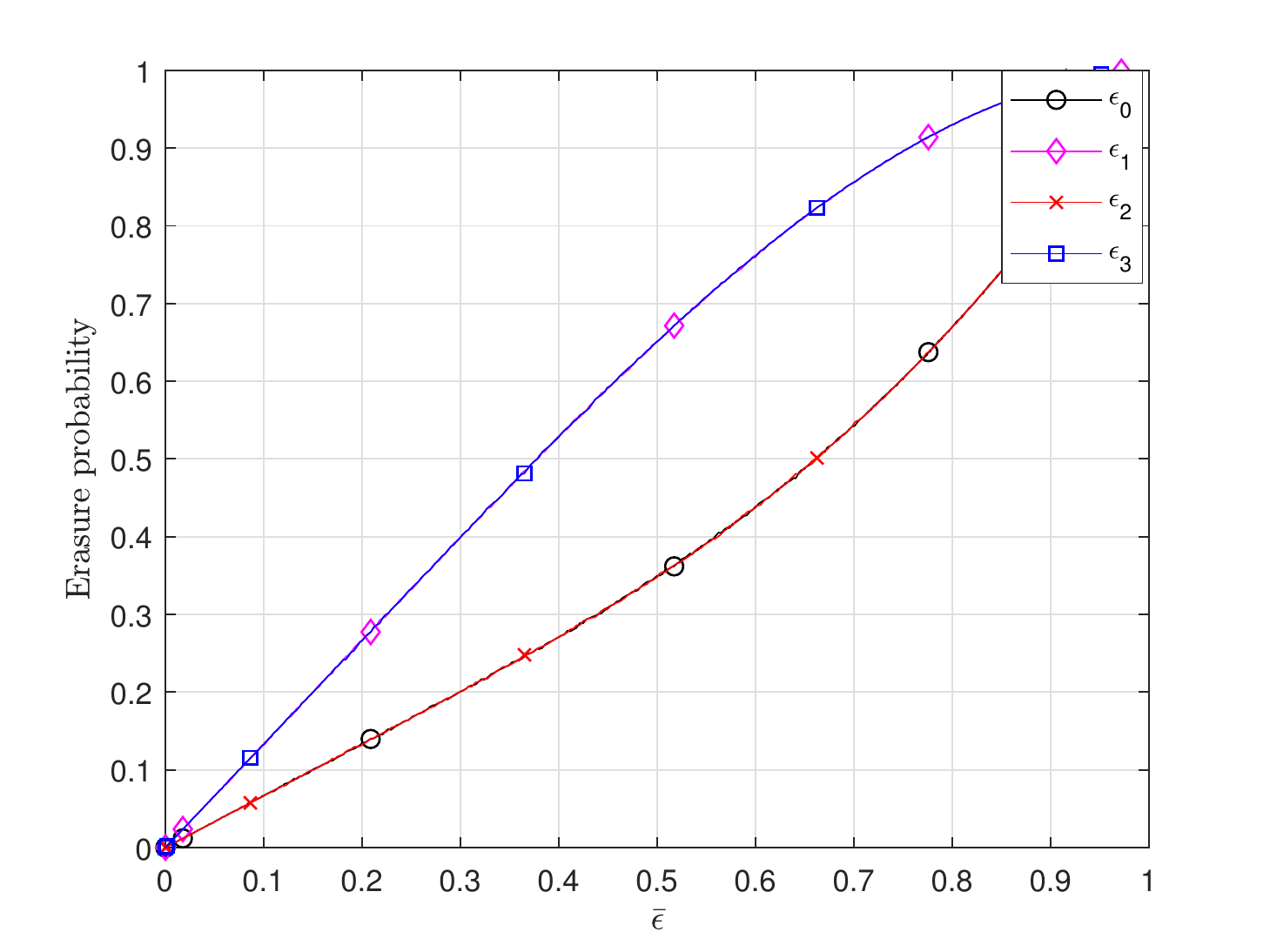}
	\vspace*{-3mm}
	\caption{Equivalent parallel BECs for uniform Gray labeled $16$-QAM BICM.~~~~~~}
	\vspace*{-3mm}
	\label{fig:eps_16}
\end{figure}

The capacity of bit-channels under the AWGN channel are converted to the corresponding parallel binary erasure channels (BECs) with equivalent channel quality. Use $\epsilon_i = 1-C_i$, to denote the erasure probability of the $i$-th BEC. And the overall quality of these $m$ parallel bit-channels is
\begin{equation}
\bar{\epsilon} = \frac{\sum_{i=0}^{m-1}\epsilon_i}{m},\quad i\in\{0,\cdots,m-1\}.
\end{equation}
In Fig. \ref{fig:eps_16}, we show the correspondence between $\bar{\epsilon}$ and $\epsilon_i$, for a 16-ary quadrature amplitude modulation (QAM). We emphasize that density evolution under the equivalent parallel BECs is effective in estimating the decoding threshold for BICM under AWGN channels. The mutual information of the channel dominates the decoding threshold for BICM schemes \cite{5205587}. Compared to the model which considers parallel binary input AWGN channels, e.g., \cite{5729846}, the complexity of the decoding threshold analysis using density evolution is greatly simplified for considering their corresponding BECs \cite{6883627}.

\section{Background on SC-LDPC codes} \label{sec:Code_cons}
This section describes SC-LDPC code ensembles in terms of their protograph representations \cite{thrope}. A protograph is a small bipartite graph, where each node and edge in a protograph represents a set of nodes and edges, with possible edge permutations, in its equivalent Tanner graph. Furthermore, an SC-LDPC code is obtained by lifting an SC-LDPC protograph. In particular, both the single-chain ensembles and the existing connected-chain ensembles are reviewed.



\subsection{Single-chain Ensemble}
\begin{figure}[h]
	\centering
	\subfigure[~~~~~~]
	{
		\includegraphics[width=0.47\textwidth, height=0.25\linewidth]{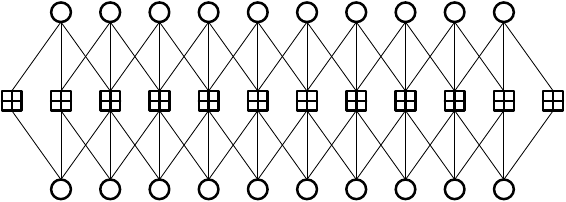}
		\label{fig:single_chain_full}
	}
\hspace*{-8mm}
	\subfigure[~~~~~~]
	{
		\includegraphics[width=0.41\textwidth]{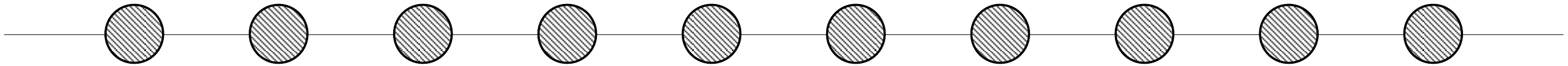}
		\label{fig:single_chain_simple}
	}
	\caption{(a) Protograph representation for ensemble $\mathcal{C}(3,6,10,2)$, (b) A simplified representation for ensemble $\mathcal{C}(3,6,10,2)$.~~~~~~}
	\label{fig:single_chain}
\end{figure}
We denote a single-chain SC-LDPC protograph ensemble by $\mathcal{C}(J,K,L,w)$. Here, $J$ and $K$ represent the variable node (VN) degree and the check node (CN) degree for the underlying regular LDPC-BC protograph, respectively. The underlying regular LDPC-BC protograph contains $b_c$ CNs and $b_v$ VNs. Furthermore, $L$ and $w$ stand for the coupling length and the coupling width, distinctively. An example of ensemble $\mathcal{C}(3,6,10,2)$ is shown in Fig. \ref{fig:single_chain_full}, where we use circles and boxes to represent VNs and CNs, respectively. Note that the CNs at the start and the end of the chain are only connected to either 2 or 4 VNs. The code rate $R$ of ensemble $\mathcal{C}(J,K,L,w)$ is given by \cite{7152893}
\begin{equation}\label{eq:rate}
R = 1-\left(\frac{(L+w) b_c}{L b_v}\right).
\end{equation}


Fig. \ref{fig:single_chain_simple} is a simplified illustration of the protograph in Fig. \ref{fig:single_chain_full} where each node illustrates a segment consisting of a CN and two VNs. Furthermore, the CNs with degree two at the termination ends are omitted in the simplified representation, leaving the boundary chain-positions with an open-end edge. 

\subsection{Connected-chain Ensembles}
In the following, two $\mathcal{C}(J,K,L,w)$ chains are connected to form a connected-chain ensemble whose code rate is the same as that of ensemble $\mathcal{C}(J,K,L,w)$ as given in Eq. (\ref{eq:rate}).

\begin{figure}[h]
	\centering
	\subfigure[~~~~~~]
	{
		\includegraphics[width=0.45\textwidth]{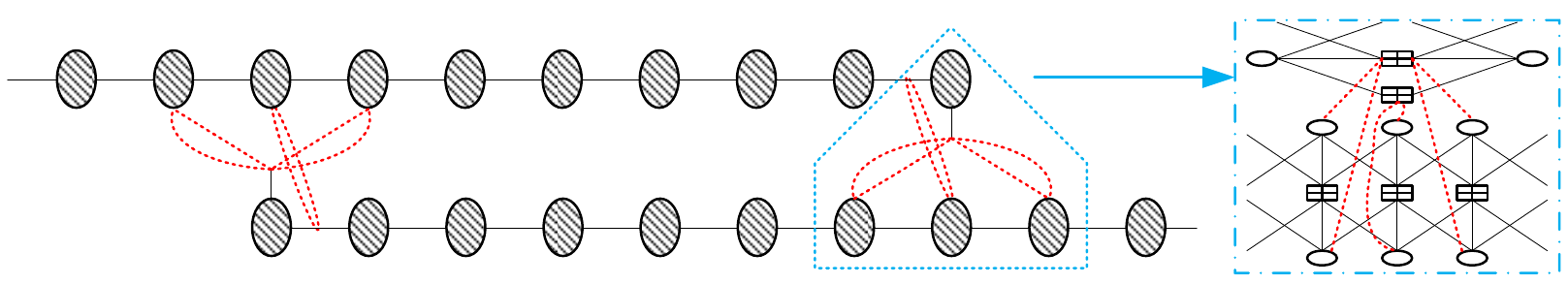}
		\label{fig:conn_chain_m2}
	}
	\subfigure[~~~~~~]
	{
		\includegraphics[width=0.45\textwidth]{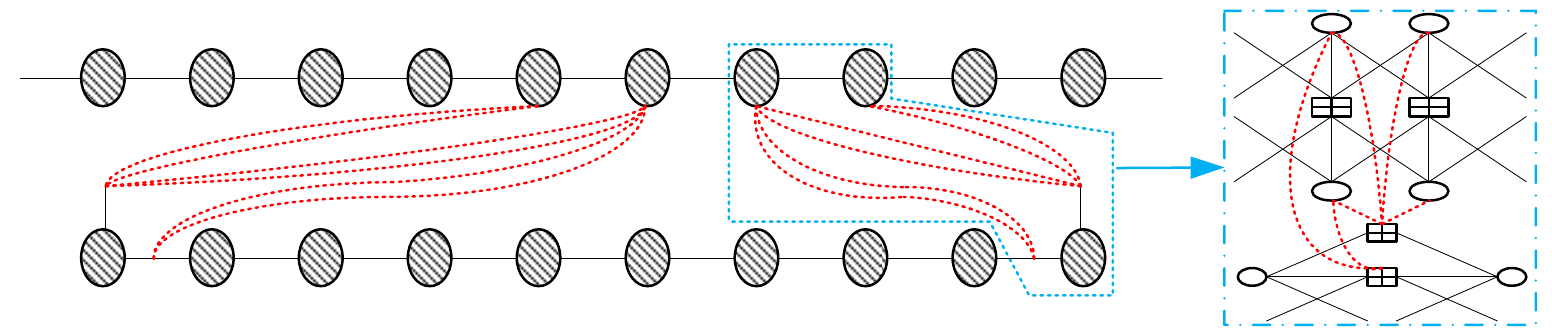}
		\label{fig:conn_chain_m1}
	}
	\caption{(a) Protograph representation for $\mathcal{L}_1(3,6,10,2)$ following \cite{6364482,6181771,8718022}, (b) Protograph representation for $\mathcal{L}_2(3,6,10,2)$ following \cite{6955088,8003438}.~~~~~~}
	\label{fig:existing_conn_chain}
\end{figure}

Targeting at improving rate/threshold trade-offs over single-chain ensembles, the authors in \cite{6364482,6181771,8718022} proposed a loop connection method by connecting two single-chain ensembles. For example, connecting two $\mathcal{C}(3,6,10,2)$ ensembles via the loop connection, denoted by $\mathcal{L}_1(3,6,10,2)$, is shown in Fig. \ref{fig:conn_chain_m2}. The loop connection effectively makes the connect-positions exhibit a similar effect as the termination ends. Therefore, the loop connection propagates the reliable information generated from both the two termination ends and the two segments of connect-positions along the two chains, as illustrated in \cite{6364482,6181771,8718022}. Hence, $\mathcal{L}_1(3,6,10,2)$ ensemble improves the decoding threshold performance over $\mathcal{C}(3,6,10,2)$ ensembles as reported in Fig. 4 of \cite{8718022}. 

Another existing connection method called the continuous-chain connection was introduced in \cite{6955088,8003438}. Specifically, a single-chain ensemble $\mathcal{C}(3,6,10,2)$ by connecting its two termination ends to the $5$-$8$ th VNs in another chain is shown in Fig. \ref{fig:conn_chain_m1}. We denote this connected-chain ensemble by $\mathcal{L}_2(3,6,10,2)$. In this case, the VNs at the connect-positions in one of the two chains have degrees of $3$, $4$, and $5$, while the degree for all VNs on the other chain remains $3$. Therefore, one chain is more reliable than the other, which creates UEP between these two chains. Furthermore, using this type of connections, a continuous transmission scheme is proposed to enforce a dependence between consecutive independent codewords, which improves the finite length performance.

In both the loop connection and the continuous-chain connection, a termination end in a chain is connected to several consecutive positions in another chain. Furthermore, the degree of VNs within a chain-position is uniform. However, the performance of connected-chain SC-LDPC can be affected by several factors, such as the choice of chains to be connected, the connect-positions, and the connect structures. Therefore, it is necessary to jointly design the connect-position and the bit mapping for connected-chain SC-LDPC coded BICM.

\section{Joint Design of Bit Mapping and Connect-Position} \label{sec:bit_map}

In this section, to address the UEP of the investigated connected-chain SC-LDPC coded BICM, we analyze the decoding threshold by using density evolution \cite{910577} over the equivalent parallel BECs. The erasure probability of each VN is associated with the bit mapping. Then, we propose an optimization algorithm, using differential evolution, to jointly design connected-chain SC-LDPC ensembles and the bit mappers that results in a good decoding threshold.
\subsection{Density Evolution and Bit Mapping}
Although single-chain SC-LDPC code has a regular VN degree, it is possible for connected-chain SC-LDPC to have an irregular VN degree. Therefore, design bit mapping that only consider regular VN degree for each chain-position as in \cite{6883627} is no longer sufficient for connected-chain SC-LDPC. Different from \cite{6883627}, this paper follows the bit mapper defined in \cite{Hager:15} and assigns bit mapping for each VN position in a protograph, which takes into account both regular and irregular VN degree for each chain-position.

Denote the number of chains to be connected and the total number of VNs in a connected chain SC-LDPC protograph as $M$ and $V$, respectively. These $V$ VNs are mapped to $m$ bit-channels, where $V=MLb_v$. This is achieved by a bit mapping matrix $\mathbf{A}=[a_{i,j}]\in\mathbb{R}^{m\times V}$, where $a_{i,j}$ denotes the fraction of the $j$-th VN being mapped to the $i$-th BEC and it must satisfy the following constraints. 
\begin{equation}
\label{eq:aij}
\left\{
\begin{aligned}
&0\le a_{i,j}\le1, \\
&\sum_{i=0}^{m-1}a_{i,j}=1, \quad j \in \{1,2,\cdots,V\},\\
&\sum_{j=1}^{V}a_{i,j}=V/m, \quad i\in\{0,1,\cdots,m-1\}.
\end{aligned} 
\right.
\end{equation}

We use $\mathcal{A}$ to denote the set of all $\mathbf{A}$ satisfying Eq. (\ref{eq:aij}). The erasure probability for the $j$-th VN in a given connected-chain SC-LDPC protograph with bit mapping $\mathbf{A}\in\mathcal{A}$, is
\begin{equation} \label{eq:eps'}
	\epsilon'_{j} = \sum_{i=0}^{m-1} \epsilon_i a_{i,j}, \quad j \in \{1,2,\cdots,V\}.
\end{equation}

Denoting $\mathcal{V}(k)$ and $\mathcal{K}(j)$ the set of VNs and CNs connected to the $k$-th CN and the $j$-th VN, respectively. In the $\ell$-th decoding iteration, the erasure probability passed from the $k$-th CN to the $j$-th VN and the erasure probability passed from the $j$-th VN to the $k$-th CN are denoted by $q_{kj}^{(\ell)}$ and $p_{jk}^{(\ell)}$, respectively. We compute $q_{kj}^{(\ell)}$ and $p_{jk}^{(\ell)}$ via density evolution \cite{910577} as the following:
\begin{align}
	&q_{kj}^{(\ell)} = 1 - \prod_{j'\in\mathcal{V}(k)\setminus j}(1-p_{j'k}^{(\ell-1)}), \label{eq:q}\\
	&p_{jk}^{(\ell)} =	\epsilon'_j \prod_{k'\in\mathcal{K}(j)\setminus k}q_{k'j}^{(\ell)}. \label{eq:p}
\end{align}

We initialize $p_{jk}^{(0)}$ to $\epsilon'_j$ following Eq. (\ref{eq:eps'}). The erasure probability of the $j$-th VN at the $\ell$-th iteration is written as
\begin{equation}\label{eq:P}
	P_{j}^{(\ell)} = \epsilon'_j \prod_{k\in\mathcal{K}(j)}q_{kj}^{(\ell)}.
\end{equation}
The decoding threshold $\bar{\epsilon}^*$ is defined as the largest $\bar{\epsilon}$ such that $P_j^{(\infty)}=0$ for all $j\in\{1,2,\cdots,V\}$ in Eqs. (\ref{eq:q}-\ref{eq:P}).
\subsection{The Proposed Algorithm}
 We use $\mathcal{P}$ to denote the set of $\mathcal{L}(J, K, L,w)$ protographs. Finding the pair of $\mathcal{L}(J,K,L,w) \in \mathcal{P}$ and $\mathbf{A}\in\mathcal{A}$ to maximize the decoding threshold $\bar{\epsilon}^*_{\mathrm{opt}}$ can be formulated as the following problem

\begin{align}
	[\mathcal{L}^*(J,K,L,w), \mathbf{A}^*] = \underset{\substack{\mathcal{L}(J,K,L,w)\in\mathcal{P},\mathbf{A}\in\mathcal{A}}}{\arg \max} ( \bar{\epsilon}^* ).
\end{align}

However, directly finding the optimal solution to this problem is difficult. Firstly, the search space for the connected-chain ensembles is enormous as it is possible to connect multiple chains from different types of codes with various coupling length, connection structure, and connect-positions. Secondly, the computational cost related to a decoding threshold calculation is large in an optimization routine. 

To reduce the search space of the connected-chain, we introduce the following constraints when connecting multiple single-chains with ensemble $\mathcal{C}(J,K,L,w)$. 

\begin{enumerate}
	\item The maximum CN degree is set to $K$.
	\item For each single chain, the CNs at only one of its two termination ends are connected to another chain. 
	\item The connect-positions and the nodes used for connecting the other chain, are identical for each chain.
	\item The designed code rate of the connected-chain ensemble $\mathcal{L}(J,K,L,w)$ should be the same as that of the single chain ensemble $\mathcal{C}(J,K,L,w)$.
\end{enumerate}

Constraint 1 guarantees that the CNs used for connecting the other chain is as reliable as other CNs in the ensemble, except the ones at the termination ends. Constraint 2 maintains two termination ends for the connected-chain ensembles to initiate the reliable information propagation and use the rest of the termination ends to generate edges for connection, to enable information propagates between multiple chains. Furthermore, constraints 1-3 limit the number of choices for VNs and CNs in each chain to be connected to other chains. And hence, this significantly reduces the search space for connected-chain ensembles.

To reduce the computation cost in the optimization routine, we propose an optimization algorithm to jointly design bit mapping and connect-positions based on differential evolution \cite{diffevo}. We denote the output of the proposed method as $[\mathcal{L}^*(J,K,L,w),\mathbf{A}^*]$. Define $\ell_s(\bar{\epsilon})$ as the number of iterations needed for $\mathcal{L}(J,K,L,w)$ and $\mathbf{A}$ to converge at $\bar{\epsilon}$, such that $P_j^{(\ell_s(\bar{\epsilon}))}=0$ for all $j\in\{1,2,\cdots,V\}$ in Eqs. (\ref{eq:q}-\ref{eq:P}). The proposed algorithm is presented as follows.

\begin{enumerate}
	\item Initialize the average channel quality $\bar{\epsilon}$ to the decoding threshold for $\mathcal{C}(J,K,L,w)$ with a uniform bit mapping $\mathbf{A}^*=\mathbf{A}_{\mathrm{uni}}$, where $\forall a_{i,j}\in\mathbf{A}_{\mathrm{uni}}, a_{i,j}=1/m$.
	\item Find $\mathcal{L}^*(J,K,L,w)\in\mathcal{P}$, which follows constraints 1-5, and $\mathbf{A}^*\in\mathcal{A}$ such that the number of decoding iterations required for convergence is minimized for a given average channel quality $\bar{\epsilon}$, i.e.,
	\begin{align}\label{eq:opt_p}
	[\mathcal{L}^*(J,K,L,w),\mathbf{A}^*] =\underset{\substack{\mathcal{L}(J,K,L,w)\in\mathcal{P},\mathbf{A}\in\mathcal{A}}}{\arg \min} (\ell_s(\bar{\epsilon})).
	\end{align} To solve the optimization problem in Eq. (\ref{eq:opt_p}), we use differential evolution \cite{diffevo}.
	\item For the found $[\mathcal{L}^*(J,K,L,w), \mathbf{A}^*]$, compute its decoding threshold $\bar{\epsilon}^*$ by using Eqs. (\ref{eq:q}-\ref{eq:P}).
	\item If $\bar{\epsilon}^*$ is larger than $\bar{\epsilon}$, set $\bar{\epsilon} = \bar{\epsilon}^*$ and go to Step 2. Otherwise, stop and output $[\mathcal{L}^*(J,K,L,w),\mathbf{A}^*]$.
\end{enumerate}

We will show that the connected-chain SC-LDPC code and the bit mapper designed with the these constraints and the simplified optimization routine improves both the asymptotic performance and the finite-length performance over its single-chain counterpart for BICM schemes in Section \ref{sec:des} and Section \ref{sec:result}, respectively.
\subsection{Design Example} \label{sec:des}
\begin{figure}[h]
	\centering
	\includegraphics[width=0.45\textwidth]{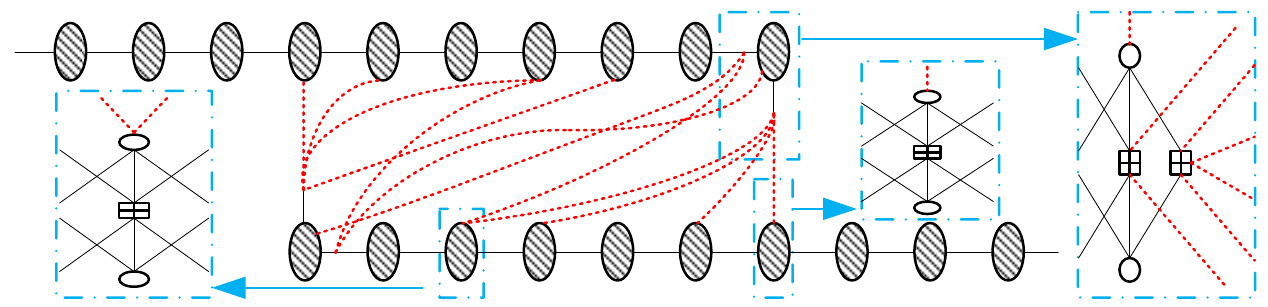}
	\caption{Protograph representation for the designed $\mathcal{L}^*(3,6,10,2)$.~~~~~~}
	\label{fig:proposed_conn_chain_conn}
\end{figure}

By considering connecting two single-chain ensembles $\mathcal{C}(3,6,10,2)$ for BICM with Gray labeled 16-QAM, we design an ensemble $\mathcal{L}^*(3,6,10,2)$ whose protograph representation and the associated bit mapping $\mathbf{A}^*$ are shown in Fig. \ref{fig:proposed_conn_chain_conn} and Table \ref{tab:bm}, respectively. In the same table, we also report the decoding threshold from the existing connected-chain ensembles $\mathcal{L}_1(3,6,10,2)$ \cite{6364482,6181771,8718022} and $\mathcal{L}_2(3,6,10,2)$ \cite{6955088,8003438} with uniform bit mapping $\mathbf{A}_{\mathrm{uni}}$, and the optimized bit mapping $\mathbf{A}^{\sim}$ by using the bit mapping optimization method in \cite{6883627, Hager:15}. As illustrated in Table \ref{tab:th}, the largest decoding threshold is obtained by the proposed connected-chain ensemble $\mathcal{L}^*(3,6,10,2)$ with its designed bit mapping $\mathbf{A}^*$. This surpasses the decoding threshold of existing connected-chain SC-LDPC codes with bit mappings designed from \cite{6883627, Hager:15} by up to $0.35$ dB. Furthermore, the decoding threshold of our designed one outperforms the single chain ensemble $\mathcal{C}(3,6,10,2)$ with the bit mapping design $\mathbf{A}^{\sim}$ by $0.80$ dB.

\begin{table*}[h]
	\centering
	\caption{The optimized bit mapping $\mathbf{A}^*$ for Gray labeled $16$-QAM BICM using $\mathcal{L}^*(3,6,10,2)$.~~~~~~}
	\label{tab:bm}
	\begin{tabular}{|c|c|c|c|c|c|c|c|c|c|c|}
		\hline
		VN position      & 1-2           & 3-4         & 5-6      & 7 & 8  & 9 & 10  & 11-12      & 13 & 14  \\ \hline
		bit-channels $\{0, 2\}$ & 0.1592      & 0.5437    & 0.9739 & 0.9920      & 0.6547    & 0.9988      & 0.6546    & 0.4495 & 0.9276      & 0.7236    \\ \hline
		bit-channels $\{1, 3\}$ & 0.8408      & 0.4563    & 0.0261 & 0.0080      & 0.3453    & 0.0012      & 0.3454    & 0.5505 & 0.0724      & 0.2764    \\ \hline
		VN position      & 15 & 16  & 17-18      & 19 & 20 & 21 & 22 & 23-24     & 25 & 26 \\ \hline
		bit-channels $\{0, 2\}$ & 0.3481      & 0.5945    & 0.3991 & 0.9268      & 0.7271    & 0.7673      & 0.9995    & 0.5239 & 0.3906      & 0.3225    \\ \hline
		bit-channels $\{1, 3\}$ & 0.6519      & 0.4055    & 0.6009 & 0.0732      & 0.2729    & 0.2327      & 0.0005    & 0.4761 & 0.6094      & 0.6775    \\ \hline
		VN position      & 27 & 28 & 29-30     & 31 & 32 & 33 & 34 & 35-36     & 37-38          & 39-40        \\ \hline
		bit-channels $\{0, 2\}$ & 0.5104      & 0.4071    & 0.5408 & 0.0244      & 0.2397    & 0.5199      & 0.4187    & 0.2914 & 0.0432      & 0.0013    \\ \hline
		bit-channels $\{1, 3\}$ & 0.4896      & 0.5929    & 0.4592 & 0.9756      & 0.7603    & 0.4801      & 0.5813    & 0.7086 & 0.9568      & 0.9987    \\ \hline
	\end{tabular}
\end{table*}
\begin{table}[h]
	\centering
	\caption{Decoding thresholds $\bar{\epsilon}^*$ for various ensembles in Gray-labeled $16$-QAM.~~~~~~}
	\label{tab:th}
	\begin{tabular}{|c|c|c|}
		\hline
		& $\bar{\epsilon}^*$ & $E_b/N_0$  \\ \hline
		$\mathcal{C}(3,6,10,2), \mathbf{A}_{\mathrm{uni}}$ & $0.5036$  &    $3.16$ dB           \\ \hline
		$\mathcal{L}_1(3,6,10,2), \mathbf{A}_{\mathrm{uni}}$ & $0.5365$  &    $2.61$ dB       \\ \hline
		$\mathcal{L}_2(3,6,10,2), \mathbf{A}_{\mathrm{uni}}$ & $0.5036$ &   $3.16$ dB        \\ \hline
		$\mathcal{C}(3,6,10,2), \mathbf{A}^{\sim}$ & $0.5187$  &    $2.91$ dB          \\ \hline
		$\mathcal{L}_1(3,6,10,2), \mathbf{A}^{\sim}$ & $0.5456$  &    $2.46$ dB       \\ \hline
		$\mathcal{L}_2(3,6,10,2), \mathbf{A}^{\sim}$ & $0.5518$ &   $2.36$ dB        \\ \hline
		$\mathcal{L}^*(3,6,10,2), \mathbf{A}^*$ & $0.5697$  &    $2.11$  dB         \\ \hline
	\end{tabular}
\end{table}

\section{Numerical results} \label{sec:result}
\begin{figure}[h]
	\centering
	\includegraphics[width=0.45\textwidth]{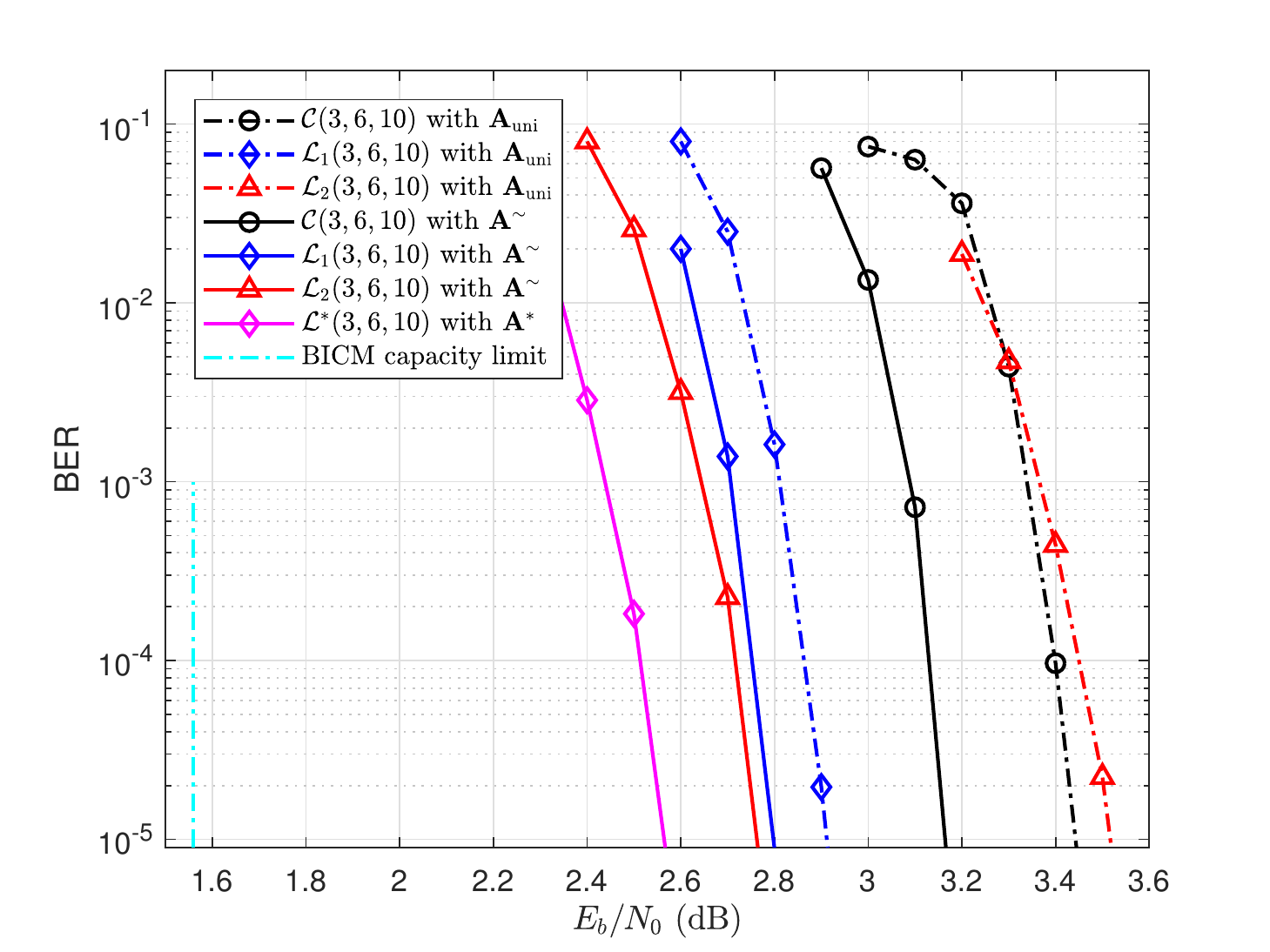}
	\caption{BER of protograph based SC-LDPC coded uniform Gray labeled $16$-QAM BICM with single-chain ensemble $\mathcal{C}(3,6,10,2)$ and connected-chain ensembles, $\mathcal{L}_1(3,6,10,2)$, $\mathcal{L}_2(3,6,10,2)$, and $\mathcal{L}^*(3,6,10,2)$, under the AWGN channel.~~~~~~}
	\label{fig:ber}
\end{figure}

In this section, we show the finite-length performance of the jointly designed connected-chain ensemble $\mathcal{L}^*(3,6,10,2)$ and the bit mapping $\mathbf{A}^*$ for Gray labeled $16$-QAM BICM under the AWGN channel. We set the code length to $N=80,000$. The protograph based SC-LDPC codes are obtained via lifting the single-chain ensemble $\mathcal{C}(3,6,10,2)$ by $4,000$ and the connected-chain ensembles, $\mathcal{L}_1(3,6,10,2)$, $\mathcal{L}_2(3,6,10,2)$, and $\mathcal{L}^*(3,6,10,2)$, by $2,000$. The investigated ensembles have a design code rate $R=0.4$. Their bit-error-rate (BER) performance is shown in Fig. \ref{fig:ber}. We also show the BICM capacity limit at a spectral efficiency of $1.6$ bits/s/Hz.


At a BER of $10^{-5}$, the jointly designed connected-chain ensemble $\mathcal{L}^*(3,6,10,2)$ and bit mapper significantly outperforms both the single-chain ensemble $\mathcal{C}(3,6,10,2)$ and the existing connected-chain ensembles, $\mathcal{L}_1(3,6,10,2)$ and $\mathcal{L}_2(3,6,10,2)$ with uniform bit mapping by $0.86$ dB, $0.35$ dB, $0.93$ dB, respectively. Furthermore, the jointly designed connected-chain SC-LDPC code and bit mapping outperforms the existing connected-chain SC-LDPC ensembles, $\mathcal{L}_1(3,6,10,2)$ and $\mathcal{L}_2(3,6,10,2)$, with the existing bit mapping design method in \cite{6883627, Hager:15} for about $0.19$ dB and $0.23$ dB, respectively. 


\section{Conclusion} \label{sec:conclus}
This paper considered the design of connected-chain SC-LDPC coded BICM. To improve the decoding threshold of connected-chain SC-LDPC coded BICM, we proposed an algorithm to jointly optimize the connect-positions with the bit mapping. Several constraints to effectively reduce the search space of connect-positions were introduced. Furthermore, we demonstrated that the designed connected-chain SC-LDPC coded BICM imposes noticeable improvements in both the asymptotic performance and finite-length performances over BICM using SC-LDPC codes constructed from the single-chain ensemble and the connected-chain ensembles \cite{6364482,6181771,8718022,6955088,8003438}.

\bibliographystyle{IEEEtran}
\bibliography{pubs}

\end{document}